\documentclass[pra,floatfix,twocolumn,superscriptaddress]{revtex4-1}
\usepackage[final]{graphicx}
\usepackage{times,bbm,amsmath,amssymb}
\usepackage{epsfig,color}
\usepackage{hyperref}
\usepackage{float,siunitx}
\usepackage[caption = false]{subfig}
\usepackage[english]{babel}
\usepackage{thumbpdf,enumerate}
\usepackage{booktabs}
\usepackage{sidecap}
\usepackage[scaled=.8]{couriers}
\usepackage{pstricks}
\usepackage{multirow}
\usepackage{placeins}
\usepackage{pst-grad}
\usepackage{epigraph}
\usepackage{longtable}
\usepackage{booktabs}
\usepackage{gensymb}

\begin{document}

\title{Quantum sensors for dynamical tracking of chemical processes}

\author{Valeria Cimini}
\author{Ilaria Gianani}\email{ilaria.gianani@uniroma3.it}
\author{Ludovica Ruggiero}
\author{Tecla Gasperi}
\author{Marco Sbroscia}
\author{Emanuele Roccia}
\author{Daniela Tofani}
\author{Fabio Bruni}
\author{Maria Antonietta Ricci}
\affiliation{Dipartimento di Scienze, Universit\`a degli Studi Roma Tre, Via della Vasca Navale 84, 00146, Rome, Italy}
\author{Marco Barbieri}
\affiliation{Dipartimento di Scienze, Universit\`a degli Studi Roma Tre, Via della Vasca Navale 84, 00146, Rome, Italy}
\affiliation{Istituto Nazionale di Ottica - CNR, Largo Enrico Fermi 6, 50125, Florence, Italy}

\begin{abstract} Quantum photonics has demonstrated its potential for enhanced sensing. Current sources of quantum light states tailored to measuring, allow to monitor phenomena evolving on time scales of the order of the second. These are characteristic of product accumulation in chemical reactions of technologically interest, in particular those involving chiral compounds. Here we adopt a quantum multiparameter approach to investigate the dynamic process of sucrose acid hydrolysis as a test bed for such applications. The estimation is made robust by monitoring different parameters at once. 
\end{abstract}

\maketitle



All modern technologies make abundant use of sensors. Their use ranges in everyday tools such as mobile technologies and GPS navigators, to industrial applications. The sought objective, common to all sensors, is to achieve the most accurate measurement of a given signal, compatibly with the  signal-to-noise ratio dictated by the quality of the hardware and the conditions of the environment. In this respect, initializing the sensor in a configuration with high sensitivity amounts to controlling its physical state. In the most general case, it is necessary to invoke the use of quantum features, in order to achieve the ultimate precision given by the Heisenberg limit \cite{giovannetti, giovannetti2}. These promises can be fully realized by means of a robust control over the quantum properties of the device, even under unfavorable circumstances.

Quantum sensing \cite{sensors1, sensors2} offers a clear framework  to study and implement measurement strategies to attain improved precision bound by exploiting the use of quantum probes.
Photonics in particular provides a flexible and versatile platform for monitoring systems: probing is performed by preparing light so to show distinctively quantum behaviors.
Phase estimation has been extensively studied \cite{paris, giovannetti3, dem} as the case in which quantum sensing with light can contribute the most.
To this aim preparation address either the quadrature of the field, or the occupation number of distinct modes \cite{pezze}.
Squeezing is a prominent example of what the control at the quantum level of quadratures gains to precision: intrinsic fluctuations of a quadrature can be reduced below the shot-noise level, translating into an improved measurement of a phase. Alternatively N00N states, quantum superpositions of N photons being in either of two modes, can offer phase precision at the Heisenberg limit. Both cases have been conceived to monitoring in faint illumination conditions. This is the case whenever the energy of the probe can damage or affect the system under investigation \cite{chiara,opti1,opti2}.  
The use of such resource states demands different measurement schemes, for instance balanced homodyne detection is very often used for squeezing, while interferometry and photon counting are preferred for N00N states. 
Following changes in the interrogated sample is one of the main purposes of sensing. When employing quantum light, the limting factor mostly comes from the detection system and from the need to accumulate enough repetitions of the measurement for a meaningful analysis. The time-scales available with photonics vary depending on wether the probe is realised with squeezed or N00N states. The former allows to reach $\mu$s time-scales and has been adopted in adaptive phase-tracking  \cite{furusawa}; the latter, within the current and near-future technology, allows to track dynamics with at best sub-second resolution: this is mostly limited by the generation rate.
It would then seem that probing dynamics with N00N states is doomed to be unsuccessful: however, some chemical reactions accumulate products over time scales fully compatible with these specifications. Among the possible chemical processes, those in which there is an alteration of the optical phase of the probe are the most intriguing, in particular those reactions leading to a change in the optical activity of the reactants and the products. Static chiral properties have already been assessed with quantum light \cite{zeilinger, optica} .

%

Monitoring the dynamics of such processes in any realistic circumstances, adds another level of complexity, as changes in the samples might not be limited to their optical activity. 
Also, the quality of the probe state itself could degrade during the time span of the reaction.  All these translate in a form of non-stationary noise polluting the measurements; due to its variations a simple pre-calibration is futile. In some instances, using a multiparameter approach provides a solution to this \cite{animesh, mihai, ol}, since it allows to estimate the parameter of interest simultaneously with the quality of the probe that investigates the sample. 

Here we report on the first application of quantum multiparameter estimation to the tracking of a chemical process, the acid hydrolysis of sucrose \cite{hydro1,hydro2,hydro3,hydro4,hydro5}, in which the optical activity of the sample undergoes a change from dexorotatory to levorotatory due to accumulation of the reaction products. Using quantum-correlated probe states, we monitor jointly the phase and the contrast of the measurement itself. This guarantees to perform an effective and unbiased estimation of the optical activity despite the unavoidable experimental instabilities due to the setup and to the sample.

Optical activity is a rotation of the polarization axis of a light wave due to the circular dichroism of chiral compounds. These molecules are not symmetric under inversion hence they address right- and left-circular polarized light differently. This property is therefore a result of the structure of the molecules themselves and is frequently manifested in biomolecules and active principles in pharmaceutics \cite{ph1,ph2}. Quantifying the optical activity offers a simple and economic way to test the correct synthesis of chiral reactions products, with respect to structure measurements when these are not strictly necessary (e.g. monitoring the correct synthesis of a known reaction, as opposed to investigating the products of an unknown one).

In this letter we consider the hydrolysis of sucrose in an acidic environment; this is a well-known reaction with no competing processes, hence it serves as an excellent test bed for our technique. In detail, sucrose is a dimer composed of glucose and fructose. When an aqueous solution of sucrose interacts with hydrochloric acid (HCl), the latter acts as a catalyst for the hydrolysis of sucrose, eventually resulting in a solution of the two monomers, glucose and fructose. Sucrose  is dexorotatory, with specific rotation $\left[\,\alpha\,\right]^{20}_D$ = 66.4$^\circ$, and so is glucose ($\left[\,\alpha\,\right]^{20}_D$ = 52.7$^\circ$). Fructose is instead levorotatory ($\left[\,\alpha\,\right]^{20}_D$ = -92.3$^\circ $), and its optical rotatory power is greater in module than glucose. The chirality of the solution, before the reaction starts, will depend only on the sucrose and will thus be dexorotatory. As the reaction is catalyzed and progresses, the solution will be a mixture of the three molecules, and its optical power will lower due to the increasing presence of fructose. When the reaction is completed all the sucrose molecules will be hydrolized and the chirality will result overall levorotatory. The time scale necessary to reach the final products will  depend on the concentration of HCl used: the higher the concentration, the quicker the reaction will be completed.

\begin{figure}[h]
\includegraphics[width=\linewidth]{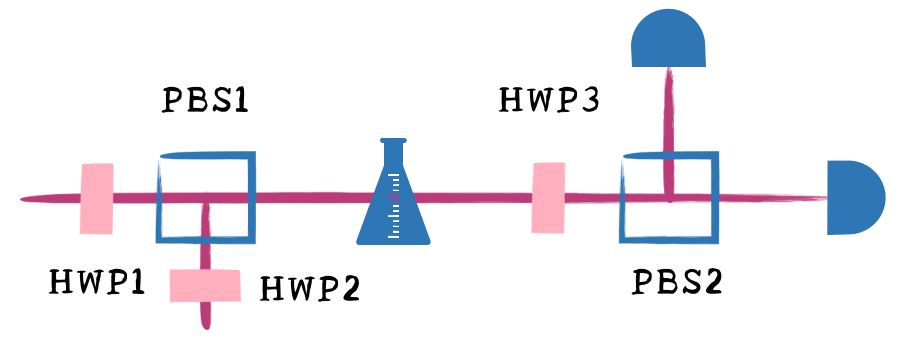}
\caption{ {\it Experimental setup.} Single photons at 810 nm, generated via type-I SPDC from a β-barium-borate (BBO, 3mm length) nonlinear crystal excited via a continuous-wave pump laser are sent through a half-wave plate (HWP1 at $0^{\circ}$ and HWP2 at $45^{\circ}$) before interfering on a beam splitter (PBS1) and the N00N state generated is sent on the chiral sample. A wave plate (HWP3) and a second polarizer (PBS2) project the outcoming photons onto different polarizations. }
\label{setup}
\end{figure}

To monitor the dynamic of the optical activity we use the multiparameter strategy proposed in \cite{optica}.  Figure \ref{setup} shows the experimental apparatus used to perform the phase estimation. Photon pairs are generated by Type-I Spontaneous Parametric Down Conversion (SPDC) and, after setting their polarization respectively to horizontal (H) and vertical (V), they are combined on a polarizing beam splitter (PBS). The two photons will have very similar spectra, hence they will be highly indistinguishable, thus their Hong-Ou-Mandel interference will result in a N00N state in the circular polarization basis,

\begin{equation}
\vert \psi \rangle= a^\dagger_H a^\dagger_V \vert 0 \rangle = \frac{1}{\sqrt{2}}(\vert 2_R,0_L\rangle + \vert 0_R, 2_L\rangle).
\end{equation}

\begin{figure}[h]
\includegraphics[width=\linewidth]{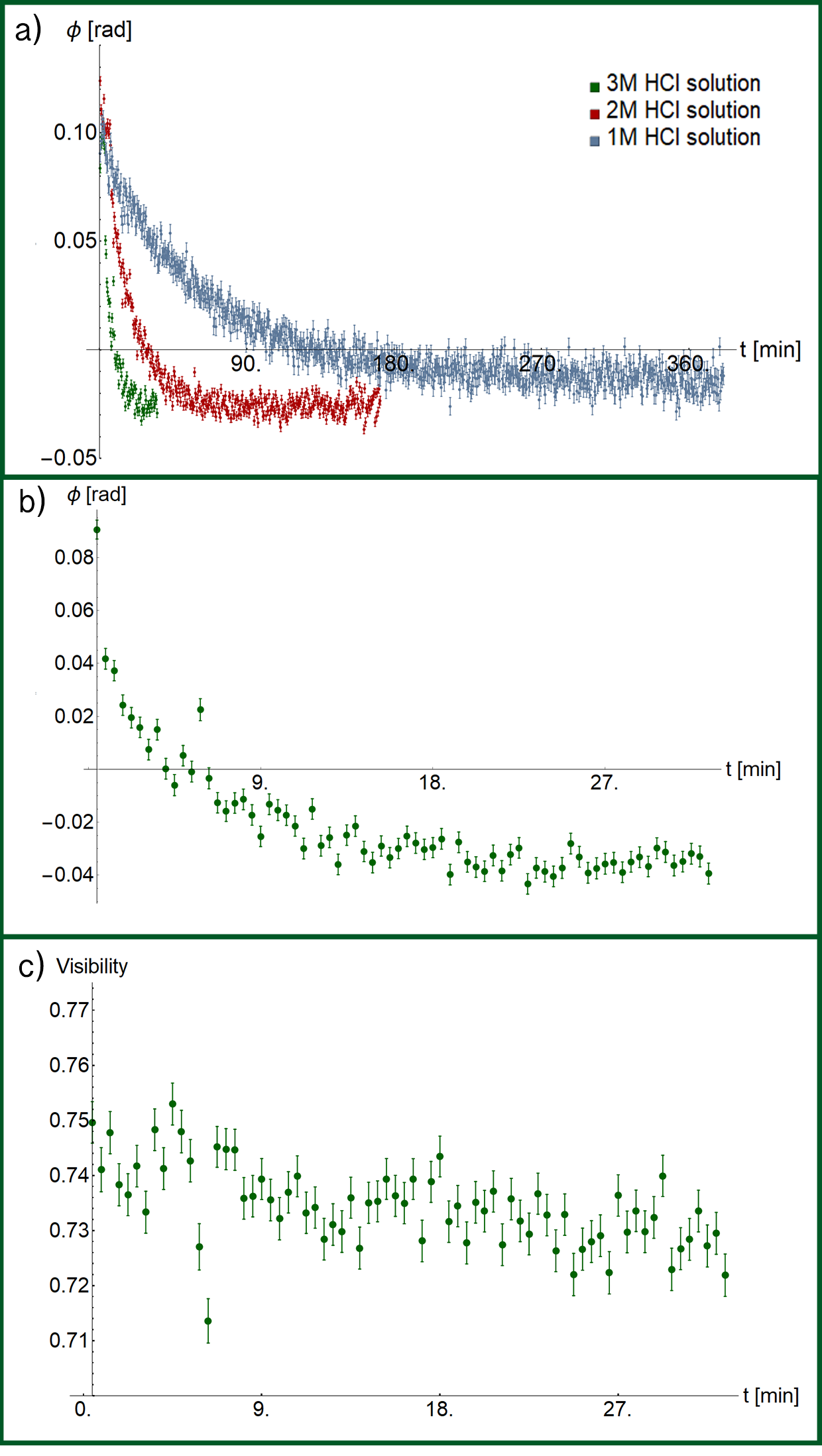}
\caption{{\it Experimental results.} a) Phase evolution for three different concentrations of HCl: 3M (dark green), 2M (teal), 1M (light green);  b) Phase behavior in function of time for the 3M HCl concentration; c) Visibility behavior in function of time for the 3M HCl concentration. }
\label{results}
\end{figure}

The chiral sample will impart a phase on the R polarization, so that after the interaction with the sample the state will read:

\begin{figure*}[t]
\includegraphics[width=\linewidth]{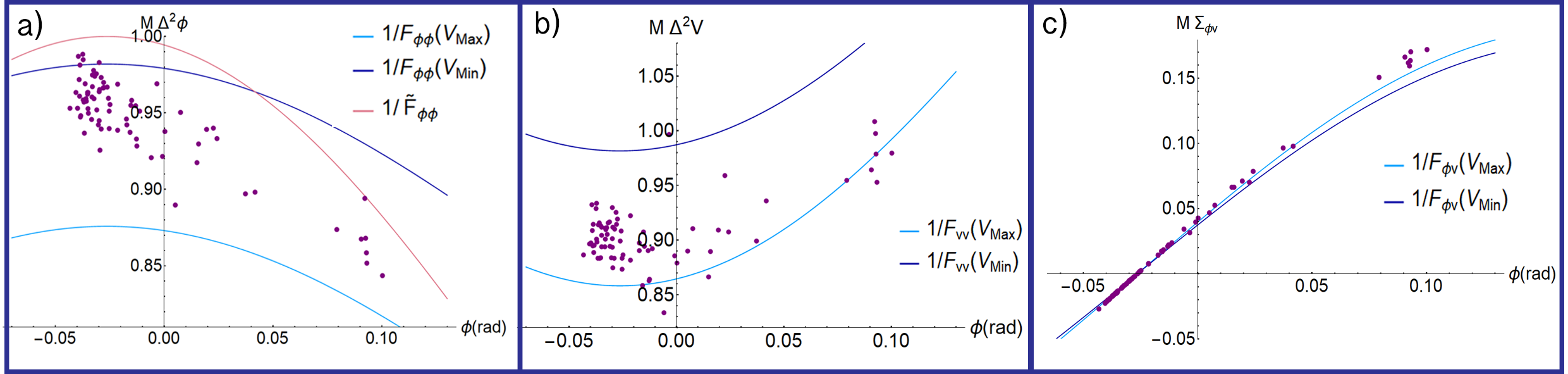}
\caption{{\it Cramer-Rao bounds.} a) Variance for the phase parameter in function of the phase itself. The curves displayed are the CRB calculated for the minimum (dark blue) an maximum (light blue) measured visibility and for the classical case (pink) b) Variance for the visibility parameter in function of the phase. The curves displayed are the CRB calculated for the minimum (dark blue) an maximum (light blue) measured visibility c) Variance for the correlations between the phase and visibility in function of the phase. The curves displayed are the CRB calculated for the minimum (dark blue) an maximum (light blue) measured visibility. $F_{ij}$ is the i,j-th element of the Fisher Information matrix.}
\label{crb}
\end{figure*}

\begin{equation}
\label{stato}
\vert \psi \rangle= \cos\phi (a^\dagger_H a^\dagger_V \vert 0 \rangle) - \sin\phi \left(\frac{(a_H^\dagger)^2-(a_V^\dagger)^2}{2}\vert 0 \rangle \right).
\end{equation} 

As the final measurement, the two optical modes are made to interfere with a controlled additional phase, and photon counting is performed  \footnote{Note that we will  post-select coincedence events. An explicit account of how this affects the acquired information content is found in \cite{optica}.}. 
However, in a realistic case, imperfections in either the state or the measurement will lead to discrepancies from the predictions obtained based only on Eq \eqref{stato}.  These would result in oscillations of the measured events depending on the imparted phase $\phi$ with visibility, $v$ equal to one. In most cases, the imperfections simply result in a reduced effective visibility, although this does not cover the whole range of alterations. Providing a simultaneous estimation for the visibility $v(t)$ and the phase $\phi(t)$ becomes of paramount importance to obtain a reliable measurement of the dynamics of the sample's optical activity when this is modified by a chemical process. Such joint estimation evades the difficulties linked to a pre-calibration of the visibility, as this might vary during the evolution. Notice that due to the spatial and time scale this method does not probe the chemical reaction itself, but rather tracks the accumulation of products; furthermore it is able to detect the formation of intermediate products. 
Multiparameter estimation is performed, at fixed time intervals, following the protocol reported in \cite{optica}, demanding to use four additional phases imparted by changing the angle $\theta$ of HWP3. The coincidence probability reads:
\begin{equation}
p(\theta \vert \phi(t), v(t)) = \frac{1}{4}\Big(1+v(t)\cos\left(8\theta-2\phi(t)\right)\Big),
\label{prob}
\end{equation}
within the generic model we employed to describe the imperfections. 

We perform the measurement on an aqueous solution of sucrose with concentration 0,3 g/mL. The sample optical path is fixed at 2 cm.  
To demonstrate that we are able to follow different dynamics in the aforementioned time scale, we report on three different data set of the same sucrose solution mixed with varying amounts of acid (1M, 2M, and 3M respectively). Typical reaction times with these acid concentrations vary between tens of minutes to a few hours.  For each measurement of $\phi (t_i), v(t_i)$, it is necessary to project the state into four different polarizations, $\theta = {0, \pi/16, \pi/8, 3\pi/16}$. Each setting of $\theta$ is acquired in 2 s. Given the time scale of the complete reaction, we perform a measurement every 30 s. 

The results for the dynamics of the sample's optical activity for the three acid concentrations are reported in panel a) of Fig. \ref{results}. These indicate that the chirality of the solution does indeed vary from dexorotatory (positive value) to levorotatory (negative value), reaching achirality at different times. We note how for the lowest acid concentration (1M) the measurement takes up to 6 hours, while for the highest concentration used(3M), the reaction is completed in approximately 30 minutes. In panels b) and c) the phase and visibility for the 3M acid concentration are presented. We observe a slight decrease in the visibility which, if not properly taken into account, would affect phase estimation, for instance introducing a bias of the estimated value. However, artefacts might still be produced whenever spurious effects differ significatively from what our model takes into account; these are reflected in the phase and visibiliity equally. 
The values of phase and visibility, as well as their uncertainties, are obtained via a Bayesian estimation based on the probability of Eq. \eqref{prob}. This procedure ensures the optimal uses of the collected count statistics. 
In order to perform the estimation it is necessary to provide a calibration value for the phase. This is obtained by performing a phase measurement with only water in the cell sample. 
The unusually low value of the visibility is to be attributed to the use of multimode fibers for the collection of the photons. These were required to account for the misalignements provoked by the sample on the photons spatial modes after mixing the solution with the HCl. 

Since our multiparameter approach grants a reliable measurement of the optical activity despite the behavior of the visibility, it is not concerning to perform the measurement without reaching high visibility values, as long as this is not detrimental to the accuracy of the measurement. In fact, while quantum correlations in the probe allow to obtain a better estimation compared to a classical probe, low visibility values can remove this advantage. The bound on the precision for each parameter is provided by the Cramer-Rao bound (CRB), which  sets a lower bound on the covariance matrix of the estimated parameters. Even in principle, it is not possible to extract all the available information on both parameters at the same time \cite{mihai, animesh,gill}: the better the bound for the phase, the worse that on the visibility. A trade-off must be chosen which amounts to investing part of the resources on estimating $v$; in our measurements, the trade-off is a function of $\phi$ only, and, for values around $\phi=0$, the resources are halved between the two estimations. 

In Fig \ref{crb} we show the CRB for the two parameters and their correlations. 
The plot in fig. \ref{crb} a) show the CRB for the phase calculated for the maximum an minimum visibility for our experiment together with the classical bound. This considers, for any value of the phase,  an ideal scenario in which multiparameter estimation is performed with classical light, given the same number of resources (i.e. detected photons), and the same trade-off between the precision. 
 For the phase values considered, even our minimum visibility still provides a better estimation than the classical case. 

Concluding, we have performed dynamical tracking of a chemical process using quantum resources.  We have demonstrated that the precision of our measurement surpasses the classical limit through the dynamics even in the presence of noise. The robustness against noise is granted by the multiparameter approach which allows to monitor the goodness of the measurement itself in real time. Our work aims at emphasizing the suitability of quantum sensing for biological and chemical uses and opens new possibilities for the application of quantum technologies in these frameworks.

%
%


\section*{ACKNOWLEDGEMENTS}

The authors would like to thank N. Spagnolo and F. Sciarrino for lending of lab equipment, and L. Mancino, M. G. Genoni, and B. Capone for stimulating discussion.


\begin{thebibliography}{99}


\bibitem{giovannetti} V.Giovannetti,S.Lloyd,andL.Maccone, Science 306, 13301336 (2004)
\bibitem{giovannetti2}. V. Giovannetti, S. Lloyd, and L. Maccone, Quantum metrology, Phys. Rev. Lett. 96, 010401 (2006).
\bibitem {sensors1} C.L. Degen, F. Reinhard, and P. Cappellaro, ”Quantum Sensing” Rev. Mod. Phys. 89, 035002 (2017).
\bibitem{sensors2} J. P. Dowling and K. P.Seshadreesan, , J. Lightwave Techno. 33, 2359-2370(2015).
\bibitem{paris} M.G.A. Paris, Int. J. Quantum Info. 7, 125-137 (2009).
\bibitem{giovannetti3} V. Giovannetti, S. Lloyd, and L. Maccone, Nat. Photonics 5, 222-229, 2011.
\bibitem{dem} R. Demkowicz-Dobrzanski, M. Jarzyna, and J. Kolodynski, Prog. Opt. 60, 345-435, 2015.


\bibitem{pezze} M. Gessner, L. Pezz´{e}, and A. Smerzi
Phys. Rev. Lett. 121, 130503, 2018
\bibitem{chiara} F. Wolfgramm, C. Vitelli, F. A. Beduini, N. Godbout, M. W. Mitchell, Nat. Photonics, 7, 28-32 (2013)


\bibitem{opti1} A Ashkin, J Dziedzic, T Yamane,  Nature, 330, pp. 769-771, 1987. 

\bibitem{opti2} N. S. Da silva and J.W. Portich, Photomedicine and Laser surgery, 28, 3, 2010.

\bibitem{furusawa}H. Yonezawa, D. Nakane, T. A. Wheatley, K. Iwasawa, S. Takeda, H. Arao, K. Ohki, K. Tsumura, D. W. Berry, T. C. Ralph, H. M. Wiseman, E. H. Huntington, A. Furusawa, Science, 337,  6101, 2012.

\bibitem{optica} E. Roccia, V. Cimini, M. Sbroscia, I. Gianani, L. Ruggiero, L. Mancino, M.G. Genoni, M.A. Ricci, and M. Barbieri,  Optica, Vol. 5 N. 10, 1171-1176, (2018).
\bibitem{zeilinger} N Tischler et al., Sci Adv., 2(10), 2016

\bibitem{animesh} M. Szczykulska, T. Baumgratz, and A. Datta, Adv. Phys. X 1, 621-639 (2016).
\bibitem{mihai}M. D. Vidrighin, G. Donati, M. G. Genoni, X.-M. Jin, W. S. Kolthammer, A. D. M. S. Kim, M. Barbieri, and I. A. Walmsley,  Nat. Commun. 5, 3532 (2014).
50.
\bibitem{ol} M. Sbroscia, I. Gianani, E. Roccia, V. Cimini, L.Mancino, P. Aloe, and M. Barbieri,  Opt. Lett. 43, 4045-4048 (2018)


\bibitem{hydro1} T. Schoebel, S. R. Tannenbaum  T. P. Labuza, Journal of Food Science, 34, 4, 1969.
\bibitem{hydro2} 
E. Tombari, G. Salvetti, C. Ferrari, and G. P. Johari, J. Phys. Chem. B, 111 (3), pp 496–501, 2007.
\bibitem{hydro3} V.H. Hartofylax, C.E. Efstathiou and T.P. Hadiioannou, Analytzca Chimzca Acta, 224, 159-168, 1989.
\bibitem{hydro4} Martínez et al., J Microb Biochem Technol, 7:5, 2015.
\bibitem{hydro5} Martínez et al. Microbial Cell Factories, 13:87, 2014.
\bibitem{ph1}W. J. Lough, I. W. Wainer, Eds., Chirality in the Natural and Applied Science, Blackwell Publishing Ltd., 2002.
\bibitem{ph2} D. B. Amabilino, Chirality at the Nanoscale: Nanoparticles, Surfaces, Materials and More, Wiley-VCH,.2009.

\bibitem{gill} R. D. Gill and S. Massar, Phys. Rev. A 61, 042312 (2000).

\end{thebibliography}
\end{document}